\documentclass[prd,11pt,a4paper,preprintnumbers]{article}
\pdfoutput=1

\usepackage{graphicx}
\usepackage{subfigure}
\usepackage{epstopdf}
\usepackage{mathrsfs}
\usepackage{amssymb}
\usepackage{verbatim}
\usepackage{color}
\usepackage{colortbl}
\usepackage{multirow}
\usepackage{amsmath}
\usepackage{amsmath}
 \usepackage{amsfonts}
 \usepackage{amssymb}
 \usepackage{mathtools}
 \usepackage{MnSymbol}
 \usepackage{multicol}
 \usepackage[colorlinks=true,linkcolor=blue,citecolor=blue]{hyperref}%
\usepackage{cite}
\usepackage{cancel}

\def\be{\begin{equation}}
\def\ee{\end{equation}}
\def\bea{\begin{eqnarray}}
\def\eea{\end{eqnarray}}

\def\comment#1{}

\newcommand{\lff}{\lambda_{\phi}}
\newcommand{\lhh}{\lambda_{h}}
\newcommand{\mff}{\mu_{\phi}}
\newcommand{\gev}{\, {\rm GeV}}

\begin{document}

\begin{flushright}
ULB-TH/16-17
\end{flushright}

\vskip 1cm

\begin{center}{\Large\bf Sneutrino driven GUT Inflation in Supergravity}\end{center}


%
\vspace{0.1cm}

\vskip 25pt

\renewcommand{\thefootnote}{\fnsymbol{footnote}}

\begin{center}
{
Tom\'as E. Gonzalo$^1$\footnote{E-mail: t.e.gonzalo@fys.uio.no},
Lucien Heurtier$^2$\footnote{E-mail: lucien.heurtier@ulb.ac.be},
Ahmad Moursy$^3$\footnote{E-mail: amoursy@zewailcity.edu.eg}
}

\vskip 10pt

{\it \small $^1$Department of Physics, University of Oslo, N-0316 Oslo, Norway}\\
{\it \small $^2$Service de Physique Th\'eorique, Universit\'e Libre de Bruxelles, Boulevard du Triomphe, CP225, 1050 Brussels, Belgium}\\
{\it \small $^3$Center for Fundamental Physics and University of Science and Technology, Zewail City of Science and Technology, $6^{\text th}$ of October City, Giza, Egypt}

\medskip

\abstract{
In this paper, we embed the model of flipped GUT sneutrino inflation --in a flipped SU(5) or SO(10) set up-- developed by Ellis et al. in a supergravity framework. The GUT symmetry is broken by a waterfall which could happen at early or late stage of the inflationary period. The full field dynamics is thus studied in detail and these two main inflationary configurations are exposed, whose cosmological predictions are both in agreement with recent astrophysical measurements. The model has an interesting feature where the inflaton has natural decay channels to the MSSM particles allowed by the GUT gauge symmetry. Hence it can account for the reheating after the inflationary epoch.
}

\end{center}

\clearpage

\renewcommand{\thefootnote}{\arabic{footnote}}
\setcounter{footnote}{0}

\section{Introduction}

In the last years, the results published by the Planck collaboration about the Cosmic Microwave Background (CMB)~\cite{Ade:2015lrj} have lead to the speculation of a connection between the ideas of cosmological inflation and supersymmetric grand unification. Using the measured value for the amplitude of scalar perturbations in the CMB, $A_s = (2.19 \pm 0.11) \times 10^{-9}$, it is possible to estimate the energy density during the inflationary epoch as
\begin{equation}
 V = (2 \times 10^{16} \gev)^4  \left( \frac{r}{0.15} \right),
 \label{inflationatunification}
\end{equation}
where $r$ is the ratio of tensor to scalar perturbations. Therefore, for a value of $r \sim 0.1$, perfectly compatible with Planck's observations, equation \eqref{inflationatunification} shows the remarkable coincidence of the energy density during inflation $V^{1/4}$ and the unification scale predicted by Supersymmetric GUTs, $M_{GUT} \sim 2\times 10^{16}$ GeV. Though only an apparent relation, since there is a priori no connection between the involved parameters, it serves as a motivational factor to build models of GUT inflation.

The issue of inflationary Grand Unification has been studied extensively in the past \cite{Lyth:1998xn,Kyae:2005nv,Antusch:2010va,Rehman:2009yj,Arai:2011nq, Ellis:2014xda,Heurtier:2015ima,Kawai:2015ryj}. These models typically require an epoch of hybrid inflation \cite{Linde:1993cn,Antusch:2008pn,Antusch:2009ef, Clesse:2010iz,Buchmuller:2014epa}, during which the inflaton field(s), as it rolls down the slope of its potential, destabilizes the GUT preserving minimum and thus breaks the symmetry. After dropping down to its new vacuum, the GUT Higgs field backlashes into the inflationay potential, causes the end of the inflationary phase.

Many of the SUSY GUTs in the literature, built to accommodate low-energy phenomena \cite{Ellis:1981tv,Ellis:1990wk,Antoniadis:1987dx} cannot be consistently constructed within a hybrid inflationary framework. This is due to the known GUT monopole problem~\cite{'tHooft:1974qc, Polyakov:1974ek}, which refers to the overproduction of magnetic monopoles during the GUT epoch, in tension with experimental searches, and having the unfortunate side effect of overclosing the universe~\cite{Guth:1980zm}. This issue is then avoided if the GUT symmetry is broken before the inflationary era so that these monopoles, along with other topological defects such as domain walls or cosmic strings, are diluted away by the rapid expansion of the universe.

However, hybrid inflation expects the unification symmetry to be broken towards the end of inflation, with potentially not enough e-folds left over to wash away the topological defects. Fortunately, as pointed out by 't Hooft \cite{'tHooft:1974qc}, monopoles do not arise in systems with non-semisimple symmetry groups, i.e. Lie-groups of the form $\mathcal{G}\times U(1)_X$, where the charge of the abelian factor $Q_X$ enters in the linear combination of the electromagnetic charge
\be
 Q_{em} = \alpha_1 Q_1 + \dots \alpha_n Q_n + \beta Q_X,
\ee
where the charge $Q_i$ and parameters $\alpha_i$ and $\beta$ ($i=1,\dots,n$) depend on the structure of $\mathcal{G}$ and the pattern of symmetry breaking.

Therefore, we will take non-semisimple groups as the unification symmetry, in particular $SU(5)\times U(1)$ and $SO(10) \times U(1)$, similar to previous works \cite{Kyae:2005nv,Rehman:2009yj,Ellis:2014xda}, where the role of the inflaton is played by the right-handed sneutrino \cite{Ellis:2014xda,Bjorkeroth:2016qsk,Peloso:2016xqq,Deen:2016zfr,Kallosh:2016sej}. We will also embed the model into a supergravity framework, with the addition of a shift symmetry\footnote{Another alternative to solve the $\eta$-problem is the addition of a Heisenberg symmetry, as discussed in \cite{Antusch:2008pn}.} which avoids the intrinsic $\eta$-problem \cite{Yamaguchi:2000vm,Brax:2005jv,Antusch:2009ef,Dudas:2014pva,Heurtier:2015ima}. Although it is not easy to build a consistent hybrid inflation model within this framework, due mostly to the chaotic nature of the inflationary potential, there are a couple of scenarios in which it can successfully be constructed, and where the model predictions for the cosmological observables lie within $2\sigma$ of the experimental measurements. 

This paper is structured as follows. In section \ref{Section:TheModel} we describe in detail the model chosen, giving the details of the relevant field content in two flipped GUT scenarios, and the part of the superpotential relevant for inflation. In section \ref{sec:inflation} we perform an analysis of the inflationary trajectories and the conditions required by the model to satisfy successful symmetry breaking, with a few comments on the predictions for the cosmological observables. In section \ref{Section:Reheating} we will discuss reheating after the period of inflation and in the last section \ref{Section:Conclusions} we will conclude with some discussion about potential issues and summarize the results.

\section{The model}
\label{Section:TheModel}

\subsection{GUT structure}
\label{sec:guts}
Candidate groups of the form $\mathcal{G} \times U(1)_X$ must preserve the chiral structure of the SM, i.e. the SM fermions must live in conjugate representations of the group. This condition restricts the landscape of groups to a handful of them, e.g.  $SU(5) \times U(1)_X$ or $SO(10) \times U(1)_X$ \cite{Bertolini:2010yz}. These, so called \textit{flipped} GUT groups, have in common a non-standard embedding of SM fermions, which follows from the requirement of $Q_X$ charge assignments so that the SM hypercharge is obtained as \cite{Bertolini:2010yz,Maekawa:2003wm}
\bea
 \begin{array}{lll}
  Y &= \frac{1}{5} \left(Q_X - Q_{Y'}\right), &SU(5)\;\;\times U(1)_X, \\
  Y &= \frac{1}{20} \left( 5 Q_X - Q_Z - 4 Q_{Y'}\right), &SO(10) \times U(1)_X, 
 \end{array}
 \label{hypercharge}
\eea
where $Q_{Y'}$ is the charge associated with the first abelian factor of the broken $U(1)_{Y'} \times U(1)_X$, subalgebra of $SU(5)\times U(1)_X$ and, similarly, $Q_Z$ is the charge associated with the subalgebra $U(1)_{Y'} \times U(1)_Z \times U(1)_X$ of $SO(10)\times U(1)_X$.

\paragraph{$\mathbf{ SU(5)\times U(1)_X}$}~

Flipped $SU(5) \times U(1)$ has been thoroughly studied in the literature \cite{DeRujula:1980qc, Barr:1981qv, Masiero:1982fe, Derendinger:1983aj, Antoniadis:1987dx, Ellis:1988tx,Kyae:2005nv}, and it is one of the preferred candidates for a pseudo-unified group, for it solves or avoids many of the everpresent problems in GUT model building. It predicts the unification of the gauge couplings $\alpha_2$ and $\alpha_3$ at $M_{GUT} \gtrsim 10^{16}$ GeV, which allows for long-lived protons \cite{Rehman:2009yj}; it provides a natural solution to the doublet-triplet splitting via the missing partner mechanism \cite{Masiero:1982fe,Kyae:2005nv}; and it can easily be extended, via the addition of sterile neutrinos, to accommodate neutrino masses \cite{Ellis:1988tx,Ellis:2011es}.

The field content of the flipped $SU(5)$ group, following the steps of \cite{Ellis:2014xda}, can be sketched as follows :
\begin{itemize}
\item The standard model (SM) matter content is contained in representations $\mathbf{10}_F$, $\mathbf{\bar 5}_F$ and $\mathbf{1}_F$, whose respective $U(1)_X$ charges are 1, -3 and 5.

\item The Brout-Englert-Higgs bosons enacting electroweak symmetry breaking are contained in $\mathbf{5}_{H_u}$ and $\mathbf{\bar 5}_{H_d}$.

\item The model is complimented with a singlet $\mathbf{1}_{S}$, necessary to provide the mixing $\mathbf{5}_{H_u} \mathbf{\bar 5}_{H_d}$ required for electroweak symmetry breaking.

\item The heavy scalars $\Sigma$ and $\bar \Sigma$, triggering the breaking of the flipped $SU(5)$ to the standard model gauge group are contained in representations $\mathbf{10}_{H}$ and $\mathbf{\overline{10}}_{H}$.

\item An additional superfield, in the conjugate representation of the 10-dimensional matter multiplet $\mathbf{\overline{10}}_F$ with $U(1)_X$ charge -1, is added to allow the introduction of a shift symmetry in the K\"ahler potential, which will be described below.
\end{itemize}

With these field representations, and taking canonic hypercharge normalization through eq. \eqref{hypercharge}, the generator $Q_{Y'}$ can be written as 
\be 
Q_{Y'} =\tfrac{1}{6}{\text diag} \left( -2,-2,-2,3,3 \right)\,,
\ee

In addition to this field content, we add $Z_2$ matter parity so as to forbid undesirable RP-violating couplings~\cite{Kyae:2005nv}. The $U(1)_X$ and $Z_2$ charges of all the involved fields, can be seen in Table \ref{tab:U1-Z2-charges}.

\begin{table}[h]
 \centering
 \begin{tabular}{c | c c c c c c c c c c}
  \hline \hline
   &$\mathbf{5}_{H_u}$ & $\mathbf{\bar 5}_{H_d}$ & $\mathbf{\bar 5}_{F}$ & $\mathbf{10}_{H}$ & $\mathbf{\overline{10}}_{H}$ & $\mathbf{10}_{F}$ & $\mathbf{\overline{10}}_{F}$ & $\mathbf{1}_{F}$ & $\mathbf{1}_S$ \\
   \hline 
   $U(1)_{X}$ & +2 & -2 & -3 & 1 & -1 & 1 & -1 & 5 & 0\\
   \hline 
   $Z_{2}$ & + & + & - & + & + & - & - & - & +\\
   \hline  \hline
  \end{tabular}
 \caption{{\footnotesize $U(1)_X$ and $Z_2$ charges of the different representations of flipped $SU(5)$ gauge symmetry.}}
 \label{tab:U1-Z2-charges}
\end{table}

In this scenario the inflaton can be taken to be the singlet $\mathbf{1}_S$ or the right handed sneutrino, $N^c$ ($\bar N^c$), embedded in the representation $\mathbf{10}_F$ ($\mathbf{\overline{10}}_F$), both of which were covered in \cite{Ellis:2014xda}. As was mentioned in the introduction, in this work we will focus on the latter case, because it presents a more interesting case due to the constraint nature of the problem.

\paragraph{$\mathbf{SO(10)\times U(1)_X}$}~

The minimal flipped $SO(10)$ model \cite{Bertolini:2010yz, Maekawa:2003wm}, though not as studied as the flipped $SU(5)$ model, still presents a very appealing candidate as an unified model. In addition to the advantages mentioned above for the $SU(5)$ case, which this model shares, an $SO(10) \times U(1)_X$ model provides a natural link with ultraviolet completions, such as heterotic string theories, for it does not require high dimensional representations which could not be obtained via a manifold compatification of string theory~\cite{Nilles:1983ge,Brignole:1997dp}.

The field content of this theory can be summarized as follows:

\begin{itemize}
 \item All the SM matter content is embedded in the representations $\mathbf{16}_F$, $\mathbf{10}_F$ and $\mathbf{1}_F$, which also contain the two MSSM Higgs bosons, two coloured Higgs fields and an additional SM singlet.
 \item Typically, one needs two pairs of symmetry breaking fields, $\Sigma$, $\bar\Sigma$ and $\Sigma'$, $\bar\Sigma'$ in the representations $\mathbf{16}_H$, $\mathbf{\overline{16}}_H$ and $\mathbf{16}_H'$, $\mathbf{\overline{16}}_H'$, respectively. One pair would break the symmetry to the flipped $SU(5)$ model and the other, containing the fields $\mathbf{10}_H$, $\mathbf{\overline{10}}_H$ from above, will continue the symmetry breaking all the way to the low energies.
 \item Again we find the need to add an additional $\mathbf{\overline{16}}_F$ representation, to realize the shift symmetry that will be described below.
\end{itemize}

\noindent The $U(1)_X$ charges of all these fields, as well as their $Z_2$ parities can be found in Table \ref{tab:U1-Z2-charges_2}.

\begin{table}[h]
 \centering
 \begin{tabular}{c | c c c c c c c c c}
  \hline \hline
   &$\mathbf{16}_F$ & $\mathbf{\overline{16}}_F$ & $\mathbf{10}_F$ & $\mathbf{1}_F$ & $\mathbf{16}_{H}$ & $\mathbf{\overline{16}}_{H}$ & $\mathbf{16}_{H}'$ & $\mathbf{\overline{16}}_{H}'$  \\
   \hline 
   $U(1)_{X}$ & +1 & -1 & -2 & 4 & 1 & -1 & 1 & -1\\
   \hline 
   $Z_{2}$ & - & - & - & - & + & + & + & +\\
   \hline  \hline
  \end{tabular}
 \caption{{\footnotesize $U(1)_X$ and $Z_2$ charges of the different representations of flipped $SO(10)$ gauge symmetry.}}
 \label{tab:U1-Z2-charges_2}
\end{table}

It is worth noting that the flipped $SO(10)$ model provides the possibility of implementing a missing partner mechanism, analogous to the case of flipped $SU(5)$, via non-renormalizable operators of the type \cite{Bertolini:2010yz, Maekawa:2003wm}
\be
 W \supset \mathbf{\overline{16}}_H' \mathbf{16}_H \mathbf{\overline{16}}_F \mathbf{16}_F.
 \label{mpmechanism}
\ee

As in $SU(5)\times U(1)$, inflation can be driven by two different components of the $\mathbf{16}_F$ field, corresponding to the $SU(5)$ singlet and right handed sneutrino cases from above, and again we will focus henceforth on the latter case.

Both of these models have identical inflationary dynamics as well as any relevant phenomena, such as reheating as will be seen later in section \ref{Section:Reheating}. This is due to the intermediate symmetry breaking step $SO(10)\times U(1) \to SU(5)\times U(1)$, happening before or simultaneously to the symmetry breaking during the waterfall. Differences between the two models lie before the inflationary epoch, as well as in sectors not relevant to inflation, where the $SO(10)\times U(1)$ model might predict additional superpotential terms not present in the minimal $SU(5)\times U(1)$ model presented above.

\subsection{Superpotential}

For either of the flipped scenarios mentioned in section \ref{sec:guts}, one can construct a superpotential containing a piece relevant for inflation given by\footnote{The $Z_2$ charges given in Table~\ref{tab:U1-Z2-charges} disallow terms that mix $\phi_i$ with $h_j$.}
\begin{equation}
W \supset S (\lff\ \phi_1 \phi_2 + \lhh\  h_1 h_2 - M^2)+\mff \phi_1 \phi_2\,,
\label{superpotential}
\end{equation}
where we have rewritten the complex fields as $\phi_{1(2)}\equiv\ \stackrel{(-)}{N^c}$, $h_{1(2)}$ as the SM singlet component of $\Sigma$ ($\bar\Sigma$) and $S \equiv \mathbf{1}_S$. This superpotential is consistent with the $Z_2$ matter parity for both types of models, but can only be generated at the renormalizable level for the $SU(5)\times U(1)_X$ model. For the $SO(10)\times U(1)_X$ model one needs to add non-renormalizable operators like that of eq. \eqref{mpmechanism}, with $\mathbf{16}_H \leftrightarrow \mathbf{16}_F$, which reduces to \eqref{superpotential} once $\bar\Sigma'$ acquires a v.e.v. in the $S$ direction, thus breaking the symmetry and recovering the flipped $SU(5)$ model.

Using this set up, the inflaton fields will be contained among the real components of the field $\phi_{1,2}$, whereas the field $S$ will play the role of stabilizer during inflation and $h_{1,2}$ are the waterfall fields. Here the term $\mff \phi_1 \phi_2$ is important to give the inflaton its mass at the true vacuum. This term explicitely breaks R-symmetry, thus henceforth we will assume that R-symmetry is broken in our scenario. In principle, a term $\mu_S S^2$ could be added to the superpotential, but it has no effect on the inflationary dynamics so it is omitted.

It is important to note that the inflaton fields, contrarily to many supergravity scenarios of large field inflation, carries here quantum numbers since it lives in multiplets of the unification gauge group. This situation has been studied in \cite{Heurtier:2015ima} where it was shown in particular that the shift symmetry (necessary for releasing slow roll polynomial inflation in supergravity) can be incorporated using conjugate representations of the gauge group in the following way\footnote{Note that the difference of factor of two is arbitrary but convenient for an easy definition of kinematically normalized real component fields.}
\begin{eqnarray}
\text{Singlet Case} :\phi &&\text{Multiplet Case} : \phi_{+/-}\nonumber\\
&~&\nonumber\\
K=\frac{(\phi+\bar \phi)^2}{2}~&~~\longrightarrow~~&~ K=|\phi_++\bar \phi_-|^2\,.
\end{eqnarray}

As mentioned before the shift symmetry enforces us to define another 10-dimensional fermionic multiplet which has $U(1)_X$ charge opposite to $\mathbf{10}_F$, in order to have a gauge invariant K\"ahler potential, as well as cancelling any potential $D$-term contributions to the scalar potential.
In such a set up, part of the combinations $\phi_1 \pm \bar \phi_2$ will constitute the inflaton (multi) fields, while the other component will get Hubble scale masses, being stabilized during inflation. We then write the kinetic part of the lagrangian as
\begin{equation}
K=|\phi_1+\bar \phi_2|^2+|h_1|^2+|h_2|^2+|S|^2-\eta |S|^4\,.
\end{equation}
The quartic term $\eta |S|^4$ is added to stabilize $S$ at a Hubble-scale mass during inflation in order to guarantee that it does not perturb inflation. Such term can be generated for instance through loops of heavy fermions coupling exclusively to the $S$ field \cite{Buchmuller:2014pla}.

\subsection{Supergravity vacua}


Before studying in detail the inflationary trajectory, let us mention that the true vacuum in this scenario preserves supersymmetry and is defined by
\begin{equation}
\phi_1=\phi_2=S=0\text{~~and~~} h_1 h_2 = \frac{M^2}{\lhh}.
\end{equation}
As sketched above, the fields $h_i$ will be responsible for the breaking of the $SU(5)\times U(1)$ at the end of inflation. The scale of such breaking is expected to happen at the Grand Unification scale, therefore we will, from now on, impose that
\begin{equation}\label{eq:GUTbreaking}
\frac{M^2}{\lhh}=M_{GUT}^2
\end{equation}
which will unambiguously fix the parameter $\lhh$ in what follows, the mass term $M$ remaining a priori free.

The breaking of the $U(1)$ symmetry will naturally generate a Goldstone mode which one can identify by performing the following field redefinitions, consistent with kinematic normalization
\begin{eqnarray}
\phi_1+\bar \phi_2 &=& \alpha_1 + i \beta_1\,,\qquad \phi_1-\bar \phi_2 = \alpha_2 + i \beta_2\,,
\end{eqnarray}
\begin{eqnarray}
S &=& \frac{s + i \sigma}{\sqrt{2}}\,,\qquad h_{1,2} = \frac{H \pm h}{\sqrt{2}}\,,
\end{eqnarray}
and finally
\begin{eqnarray}
h &=& \frac{h_r + i h_i}{\sqrt{2}}\,,\qquad H = \rho \exp\left(\frac{i}{\sqrt{2} }\frac{\theta}{M_{GUT}}\right)\,.
\end{eqnarray}
With these definitions, $\theta$ is exactly massless and is the Goldstone boson of the theory. This will be true both during and after inflation, and as described in \cite{Heurtier:2015ima} the field $\theta$ can be removed from the theory by a simple gauge transformation. Hence, in the vacuum the remaining fields, after elimination of the goldstone mode, get masses
\begin{eqnarray}
m_s^2&=&m_{\sigma}^2 \approx M^2\lhh\,,\nonumber\\
m_{\alpha_1}^2&=&m_{\beta_1}^2=m_{\alpha_2}^2=m_{\beta_2}^2\approx \mff^2\,,\\
m_{h_r}&\approx &2M^2\lhh \,,~~ m_{h_i}^2=\frac{4g^2 M^2}{\lhh}\,,~~\text{and~~}m_{\rho}^2\approx 4 M^2\lhh\,.\nonumber
\end{eqnarray}

\section{Inflationary trajectory}
\label{sec:inflation}

We will now enter the description of the inflationary trajectory, ensuring that all directions, except the inflaton one, are prevented to run away during the Hubble scale inflation period. One also has to control that spectator fields acquire Hubble scale masses during inflation, such that their dynamics does not interfere with the inflation slow rolling.

As described above, the inflationary direction is triggered, due to the shift symmetry introduced, by the combination $\phi_1-\overline{\phi_2}\equiv \alpha_2+i \beta_2$. At very large values of the latter, all the spectator fields $S$, $h$, $\alpha_1$, $\beta_1$, as well as the heavy scalar $\rho$ ($= H$) are stabilized close to the origin, while the inflaton pair of fields $(\alpha_2, \beta_2)$ rolls down the potential. At some critical value of the inflaton, the field $\rho$ becomes tachyonic at the origin and a waterfall happens, ending in the SUSY vacuum described above, and spontaneous breaking of the unification gauge group occurs.

At early stages of inflation -- far before the waterfall happened -- the vacuum expectation values (vev) of the spectator fields are given by
\begin{eqnarray}
\langle \sigma\rangle &=&\langle \alpha_1\rangle=\langle \beta_1\rangle=\langle h_r\rangle=\langle h_i\rangle=0
\end{eqnarray}
and
\begin{equation}
\langle s\rangle = \frac{\sqrt{2}\ \lff\ \mff}{2 \eta  \lff^2-\mff^2}\ +\ \dots \,,
\end{equation}
where we expanded for large values of the inflaton fields $(\alpha_2, \beta_2)$. The masses of these fields are then at the Hubble scale,
\begin{eqnarray}
m_s^2&\approx & m_{\sigma}^2\approx 12 \eta H^2 \,,\quad m_{\rho}^2\approx  6 H^2\,,
\label{mass1}
\end{eqnarray}
\begin{eqnarray}
m_{\alpha_1}^2&\approx & m_{\beta_1}^2\approx 6  H^2\,,\quad m_{h_r}^2\approx  m_{h_i}^2\approx 3  H^2\,,
\label{mass2}
\end{eqnarray}
where $H^2$ is given by the (early)inflationary potential
\begin{equation}
H^2 = \frac{V(\alpha_2, \beta_2)}{3}\approx \frac{1}{48} \left(\alpha_2^2+\beta_2^2\right) \Big[(\alpha_2^2+\beta_2^2) \left(\lff^2-3 \mff^2\right)+8 (\mff^2+\lff M^2)\Big]\,.
\label{hubble}
\end{equation}
Note that we expanded here for large values of the inflaton pair, and large parameter $\eta$ \footnote{The latter coming from integration of a UV sector at the scale $M_p\gtrsim \Lambda \gtrsim M_{GUT}$, is of order $1\lesssim\Lambda^{-2}\lesssim~100$ in Planck units, which justifies such expansion.}. Hence, for reasonable values of $\eta \gtrsim 1/12$, all the spectator fields get masses higher than the Hubble scale, ensuring that the slow rolling of the inflaton is not perturbed by light field dynamics at early stages.

\subsection{Radiative Corrections}

In most inflationary scenarios, the tree-level scalar potential in equation \eqref{hubble} gets contributions from radiative corrections, at one-loop through the Coleman-Weinberg corrections \cite{Coleman:1973jx} or at the two-loop level due to the so called Dvali problem \cite{Dvali:1995fb}.

The Coleman-Weinberg radiative correction to the scalar potential is given by \cite{Antusch:2010va}
\begin{equation}
 V_{\text{1-loop}}(\phi) = \frac{1}{64\pi^2} \text{Str} \left[m(\phi)^4 \left(\log\left(\frac{m(\phi)^2}{Q^2}\right) - \frac{3}{2}\right) \right],
\end{equation}
where the supertrace is taken over all superfields with inflaton dependent masses $m(\phi)$. As can be seen in equations \eqref{mass1} and \eqref{mass2}, the field dependent masses are of the order of $H^2 \sim ~\tfrac{m^2 \phi^2}{3 M_p^2}$, with $H^2 \sim 10^{-10} M_p$. Hence the contribution to the one loop CW goes like 
\begin{equation}
 V_{\text{1-loop}} \sim \frac{H^4}{64 \pi^2} \left( \log\left(\frac{H^2}{Q^2}\right) -\frac{3}{2}\right).
\end{equation}
Accordingly, the 1-loop correction is of order $10^{-20}$ or smaller for any relevant renormalization scale $Q$, which is negligible compared to either the quadratic or quartic terms shown in equation \eqref{hubble}.

In the context of hybrid inflation where the inflationary trajectory is flat at tree level, it has been shown that the presence of gauge fields can significantly alter the slow rolling of the inflaton field \cite{Dvali:1995fb}. Indeed, at the two loop level, the presence of the latter might induce non negligible contribution to the inflationary trajectory, leading to a gauge $\eta$-problem.
However, as detailed in \cite{Antusch:2010va}, the fact that the unification gauge symmetry is broken during inflation by the inflaton field, provides a huge mass to the gauge fields, whose presence at the two-loop order is thus suppressed by their propagators in the loops.

In our scenario, the situation is slightly different than in the case of \cite{Antusch:2010va}, in the sense that the supersymmetry breaking scale $m_{\cancel{SUSY}}$ during inflation (with which the two loop contributions will scale due to non-renormalization theorems) is mainly due to the F-terms $F_{\phi_1}$ and $F_{\phi_2}$ and not from the mass scale $M$, as one can see from numeric relations obtained in section \ref{sec:observables}.

On the other hand, the supersymmetric masses of the non-inflaton fields have been derived to be of order $\mathcal{O}(1-10) H$ where $H\sim\mu_{\phi}\Phi$ during inflation. The gauge boson masses during inflation, of order $M_g\sim g \langle \Phi \rangle$ ($g$ being the gauge coupling) thus induce mass corrections for the inflaton of possible orders
\begin{equation}
\delta m_{\Phi}^2 \quad \sim \quad \frac{g^4}{(4\pi)^4}\frac{m_{\cancel{SUSY}}^4}{M_g^2}\,,\qquad\frac{g^4}{(4\pi)^4}\frac{m_{\cancel{SUSY}}^4H}{M_g^3}\,,\qquad\frac{g^4}{(4\pi)^4}\frac{m_{\cancel{SUSY}}^4H^2}{M_g^4}\,.
\end{equation}
Assuming that $\Phi\sim M_p$, $g\sim 0.1$, $H\sim 10^{-5}M_p$ and $m_{\cancel{SUSY}}\sim F_{\phi_i}\sim \mu_{\phi}\langle\Phi\rangle\sim H$ one get the following orders of magnitude
\begin{equation}
\delta m_{\Phi}^2 \quad \sim \quad 10^{-25}M_p^2\,,\qquad\quad 10^{-30}M_p^2\,,\qquad\quad 10^{-31}M_p^2\,,
\end{equation}
which are all negligible compared to the bare inflaton mass $\mu_{\phi}\sim 10^{-6}M_p$. For larger values of the inflation field $\Phi \gtrsim M_p$, the gauge coupling $g$ becomes really small due to gravitational corrections~\cite{Robinson:2005fj}, so this argument stands. We can thus safely neglect two loops corrections in this analysis.

\subsection{Waterfall}

Looking closer at the mass of $\rho$, assuming for simplicity that $\mff\ll \lff, M$, and noticing that $\lhh=\left(\frac{M}{M_{GUT}}\right)^2\gg M^2$ (in Planck units) provides
\begin{eqnarray}
m_{\rho}^2&\approx & \frac{\lff^2}{8} \left(\alpha_2^2+\beta_2^2\right)^2 -\frac{\lff \lhh}{2} \left(\alpha_2^2+\beta_2^2\right) \,,
\end{eqnarray}
and it appears that, while the inflaton fields decrease, $m_\rho^2$ can reach negative values at a critical value
\begin{equation}\label{eq:waterfall}
\phi_c^2\equiv\left(\alpha_2^2+\beta_2^2\right)_c \approx \frac{4 \lhh}{\lff}\,.
\end{equation}
Upon reaching this point, a waterfall is expected to take place. Demanding that the waterfall happens at fields values of order $\left(\alpha_2^2+\beta_2^2\right)_c\sim 1$ would hence imply that
\begin{equation}
\lff \approx 4\lhh\,.
\label{lhlf}
\end{equation}

\subsection{Inflationary scenarios}

In most of hybrid inflation models, it is usually assumed that reaching the critical point is equivalent to the ending of the inflationary period. Though it was shown in \cite{Clesse:2010iz, Buchmuller:2014rfa} that this assumption is far from being exact. Indeed, the transverse momentum of the inflaton field is highly diluted during the inflationary phase and it can reach the critical point with almost zero transverse impulsion. Moreover, quantum excitations of the waterfall fields can make it backreact and contribute highly to the energy density, releasing most of the required inflationary e-folds in late stages of inflation. In a nutshell, inflation can actually continue (and even be entirely released in some cases) during the waterfall period itself.

In our setup the critical point can be tuned to be at very low values of the inflaton or very large ones. In principle, for an intermediate case, the inflaton could begin falling from high -- transplanckian -- values, far beyond the critical point, starting inflating the universe. Inflation could then be continued during or after the waterfall phase, the inflaton finally running down along the $\phi$-dependent vev of the $\rho$ field.

Since the waterfall phase is rather technical to describe in an analytic way, we better consider two extreme scenarios in this paper:
\begin{enumerate}
\item {\bf Small critical value: $\phi_c\lesssim 1 M_p$}

In this case there is not enough excursion between the critical point and the true vacuum in order to release a significant number of e-folds during the waterfall phase. Spontaneous breaking of the grand unification scenario takes place at the end of inflation. 

In this case the effective potential during the inflation is given by
\begin{equation}V(\alpha_2, \beta_2)= \frac{1}{16} \left(\alpha_2^2+\beta_2^2\right) \Big[(\alpha_2^2+\beta_2^2) \left(\lff^2-3 \mff^2\right)+8 (\mff^2+\lff M^2)\Big]\,.
\end{equation}
It turns out that the potential has a rotational symmetry in the plane of $\alpha_2$ and $\beta_2$ and in the polar representation it depends only on the radial part, which we denote by $\Phi = \sqrt{\alpha_2^2+\beta_2^2}$. Hence the potential reduces to the form
\begin{equation}
V(\Phi)= \frac{1}{16} \Phi^2 \Big[\Phi^2 \left(\lff^2-3 \mff^2\right)+8 (\mff^2+\lff M^2)\Big]\,.
\end{equation}

\item {\bf High critical value: $\phi_c\gtrsim 20 M_p$}

In this case the horizon crossing happens far after the waterfall. The field $\rho$ stabilizes to its field dependent vev, and the scalar potential is hence modified, but still almost quadratic. Inflation thus happens in the valley of non vanishing $\rho$. The inflaton dependent vev of $\rho$ is then very high at the beginning of inflation and fall to the GUT scale when inflation ends.

In this case there is back-reaction to the above inflationary potential and hence the potential is given by
\bea
V(\Phi)&=&\frac{1}{16 e \eta ^3 {\lambda_\phi}} \Phi^2 \Bigg[8 \eta^3 {\lambda_\phi} {\mu_\phi}^2 + 12 \eta^3 {\lambda_h} {\mu_\phi}^2+\eta^2 {\lambda_\phi} \Phi^2 \left(\eta  \left({\lambda_\phi}^2 - 3 {\mu_\phi}^2\right)-{\mu_\phi}^2\right)\\ \nonumber
&+&12 \eta^2 {\lambda_\phi} {\mu_\phi}^2 + 6 \eta ^2 {\lambda_\phi} {\mu_\phi}^2 + 6 \eta  {\lambda_\phi} {\mu_\phi}^2+ {\lambda_\phi} {\mu_\phi}^2+2 \eta  M^2 \left(4 \eta^2 {\lambda_\phi}^2+2 \eta  {\mu_\phi}^2 + {\mu_\phi}^2\right)\Bigg]\,.
\eea
\end{enumerate}

Such scenario where the symmetry is broken at (very) early stages of inflation implies that the inflaton field value may get inhomogeneous at some point, and that derivatives of the field may dominate over vacuum energy. However it was shown in \cite{Linde:1994wt} in the context of {\em new inflation} scenarios that tiny homogeneous regions of spacetime, separated by domain walls, get inflated by the energy density contained in the domain wall themselves. Noting that in our case the phase transition appears in a context of an inflating universe, we assume that once the symmetry is broken, the radii of the homogeneous patches are bigger than the Hubble scale, such that the second inflationary phase can occur after the breaking. If such assumption does not hold, the case with large field critical value may not be valid anymore in this context.

\begin{figure}
\begin{multicols}{2}
\begin{center}
$\phi_c=1 M_p$

\includegraphics[width=\linewidth]{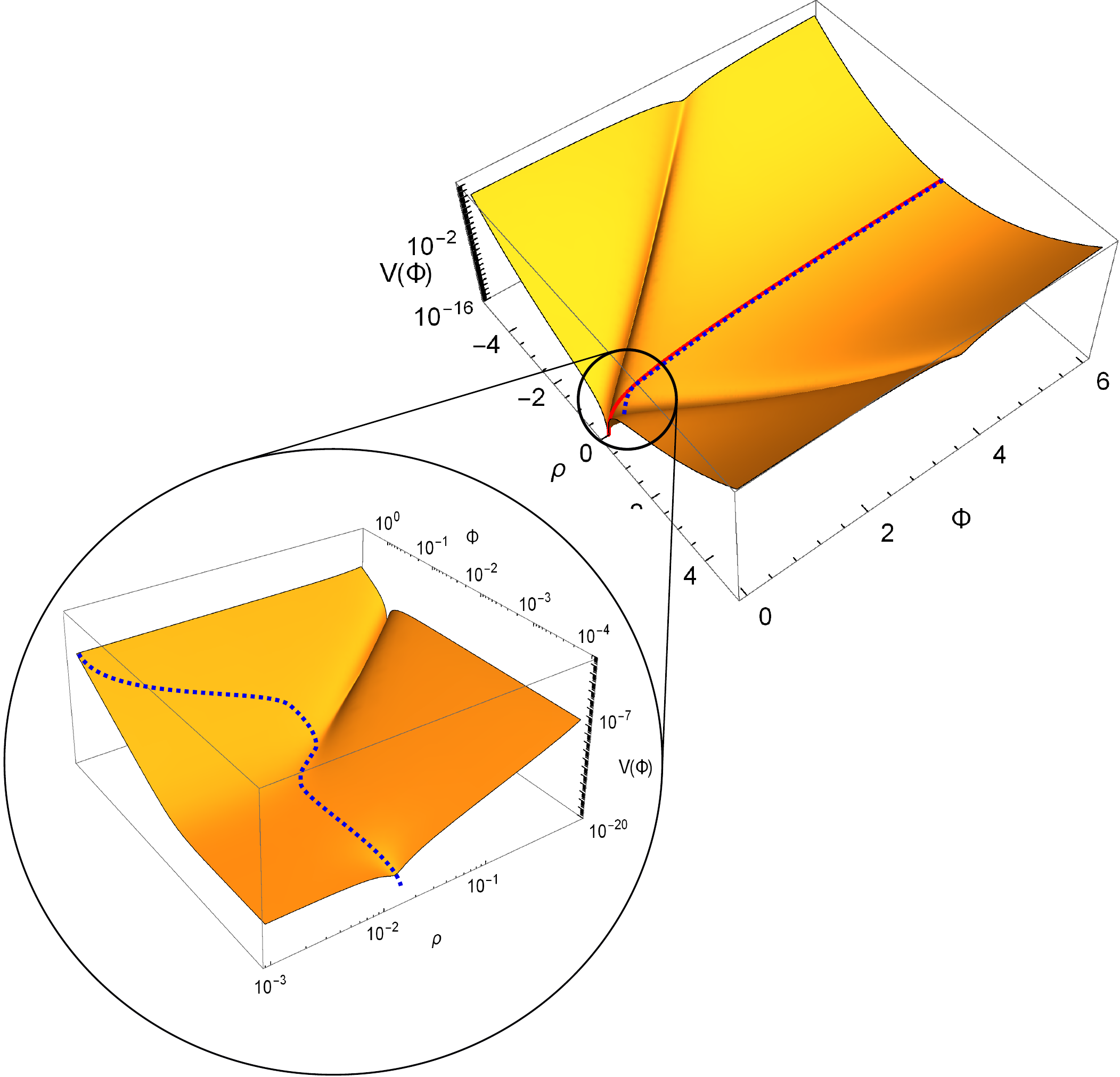}
\end{center}
\columnbreak
\begin{center}
$\phi_c=30 M_p$

\includegraphics[width=\linewidth]{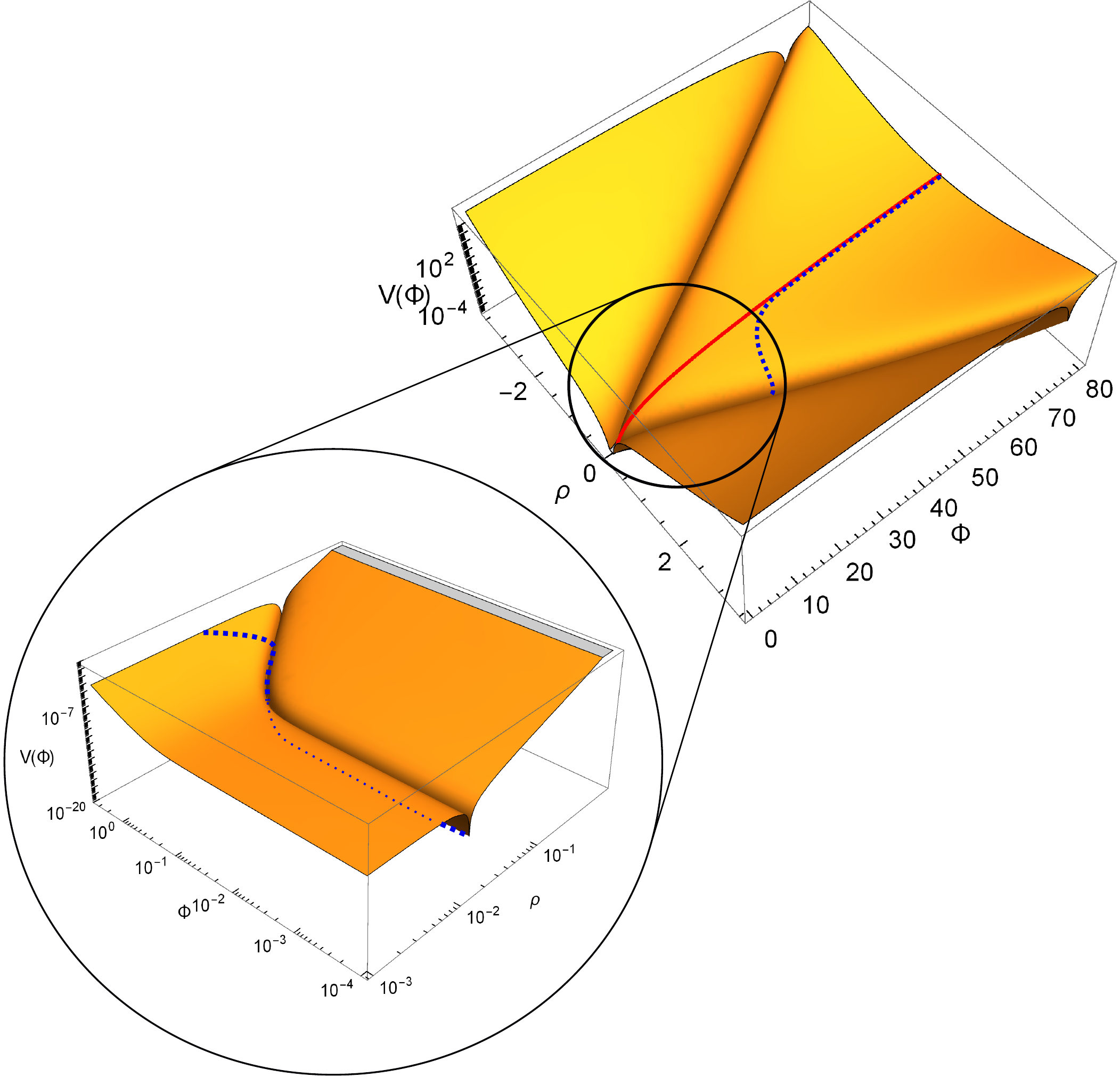}
\end{center}
\end{multicols}
\caption{\label{fig:scenarios} \footnotesize Schematic representations of the inflationary scenario, in the case where the critical point stands either at small values (left panel) or large values (right panel), corresponding respectively to the cases 1. and 2. enumerated in the text.}
\end{figure}

Both scenarios are depicted in Fig. \ref{fig:scenarios} where the inflationary trajectories are drawn for $\phi_c = 1 M_p$ and $30 M_p$.

\subsection{Observables}~
\label{sec:observables}

In both of the cases described above the quartic contribution to the potential is triggered by a term scaling like $\Phi^4(\lambda_{\phi}^2-3\mu_{\phi}^2)$. The only way inflation observables can fit with experimental constraints is to kill such a steep contribution to the scalar potential, by requiring that
\begin{equation}
\lambda_{\phi}^2= (1+\epsilon)  3 \mu_{\phi}^2\,,\qquad |\epsilon| \ll 1\,.
\label{finetuning}
\end{equation}

\begin{figure}[ht]
 \includegraphics[width=0.49\linewidth]{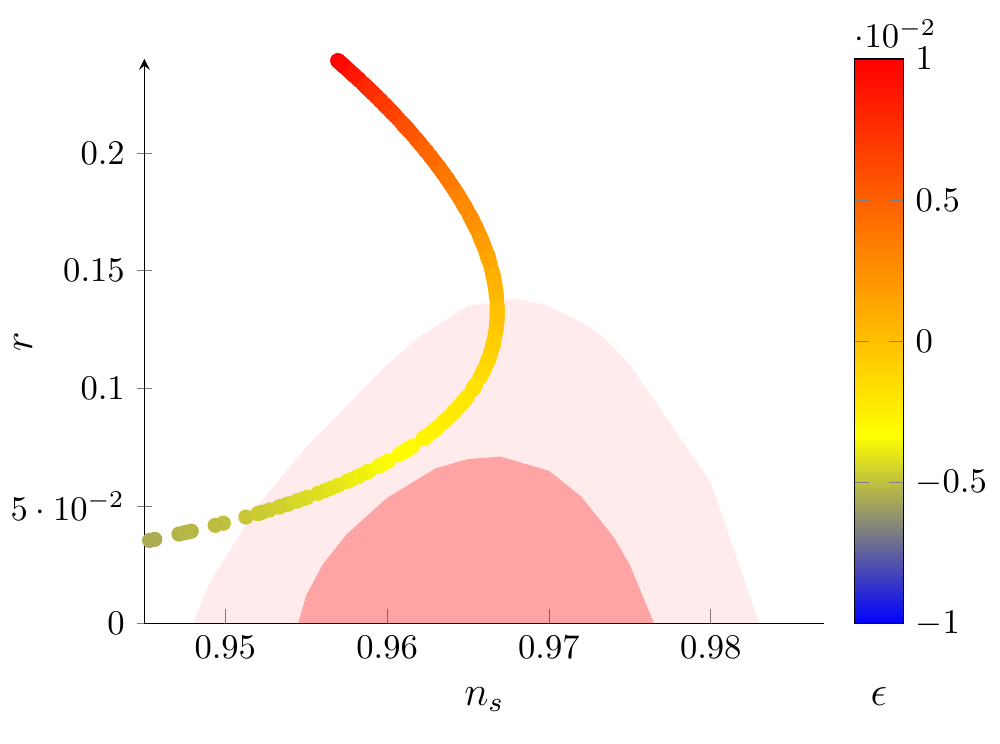}
 \includegraphics[width=0.49\linewidth]{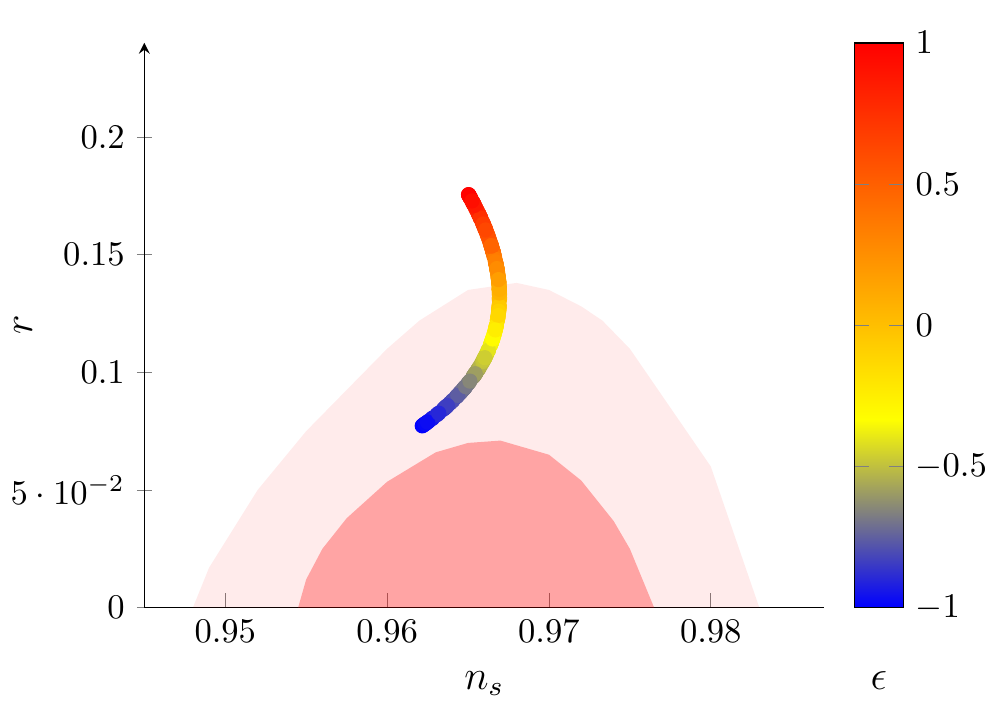}
 \caption{\footnotesize Scan over the parameters $\mu_{\phi} = [10^{10},10^{12}]$ GeV and $|\epsilon|<1$ in the space of observables in two different cases: the scenario where the critical point stands at small values ($\phi_c=1M_p$, left panel) or large values ($\phi_c=30M_p$, right panel), corresponding respectively to the cases 1. and 2. enumerated in the text. Light red regions correspond to the 1 and 2 sigma exclusion limits released by the Planck collaboration (\cite{Ade:2015xua}, {\it Planck} TT+LowP).}
 \label{fig:observables}
\end{figure}

In both scenarios, we perform a scan over the parameters $\mu_{\phi}$ and $|\epsilon| <1$. Moreover we fix $\eta$ to be of order\footnote{Note that the influence on such parameter of the observables is negligible.} $10 M_p$, while other parameters are fixed by the choice of $\phi_c$ (Eq. \eqref{eq:waterfall}) and requiring a GUT scale symmetry breaking (say $M_{GUT}=10^{-2} M_p$ in Eq. \eqref{eq:GUTbreaking}). In the case of low critical point we fix $\phi_c = \phi_{end}=1 M_p$ whereas in the case of large critical point we impose $\phi_c =30 M_p$. Such choices are made so that one can safely sit either in one or the other of the two extreme cases described above. Observables are depicted in Fig. \ref{fig:observables}.
The value of $\epsilon$ is indicated in color, in order to quantify the fine-tuning required to produce acceptable observables. It turns out that the scenario where the GUT symmetry is spontaneously broken before the horizon crossing is the less fine tuned one\footnote{Note the $10^{-2}$ difference of scale between the two plots.}. From this point of view, the choice of the flipped $SU(5)$ case as a unification gauge group becomes less necessary since magnetic monopoles generated by such phase transition will naturally be diluted by the 60 e-folds of inflation coming after all. However, this choice of gauge group has interesting consequences for reheating, since it provides natural couplings of the inflation to the MSSM particles, as described below.
%

\subsection{Effect of SUSY Breaking}


Throughout the whole analysis we have worked in a context where supersymmetry is restored after inflation.

However, in a realistic model, supersymmetry is supposed to be broken in the true vacuum~\cite{Linde:2016bcz}. It is well known that such breaking in the true vacuum, realized for instance by non vanishing F-term, has to be compensated by a constant superpotential $W_0$ in order to cancel the cosmological constant. Such constant can backreact on the inflationary trajectory, thus constraining the scale of SUSY breaking.

We will briefly estimate such constraint by investigating the effect of a simple SUSY breaking sector on our inflationary scenario. 

Let us add a Polonyi sector to our model as follows
\begin{equation}
W_{\cancel{SUSY}}= f X + W_0\,,
\label{superpotential-SUSYBreaking}
\end{equation}
\begin{equation}
K_{\cancel{SUSY}}=|X|^2 - \xi |X|^4\,,
\end{equation}
where $f$ represents the SUSY breaking scale and $W_0$ is a parameter that can be chosen such that the vacuum after inflation is Minkowskian. Accordingly, $W_0$, the gravitino mass in the vacuum, is given by
\begin{equation}
W_0 \approx \frac{f}{\sqrt{3-\frac{2M^2}{\lambda_h}}}\,,
\end{equation}
During inflation, the Polonyi field $X$ is stabilized close to the origin. The SUSY breaking scale $W_0\sim f$  back-reacts on the inflationary potential since it now enters the value of the inflaton dependant vev $\langle s\rangle$. As it is defined in this example, the SUSY breaking does not affect the rotational symmetry and one can derive the inflationary potential after integrating out the heavy fields as
\begin{eqnarray}
V(\alpha_2, \beta_2)&=& \frac{1}{16} \left(\alpha_2^2+\beta_2^2\right) \Big[(\alpha_2^2+\beta_2^2) \left(\lff^2-3 \mff^2\right)+8 (\mff(\mff+3W_0)+\lff M^2)\Big]. \nonumber \\
\end{eqnarray}
In the same approximation regime $\lambda_{\phi}^2\approx 3\mu_{\phi}^2$, in which the quartic term is subdominant, this potential can be written as
\begin{equation}
V(\Phi)\approx\frac{1}{2}\tilde m^2 \Phi^2\,,
\end{equation}
where $\Phi^2=\alpha_2^2+\beta_2^2$ and $\tilde m^2=\mff(\mff+3W_0)$.

In order to fit with experimental data, $\tilde m$ should be of order $10^{-5}M_p$. This constraint can be seen as two fold. If the SUSY breaking scale is comparable to the scale $\mff$, such constraint provides a one to one relation between the two, and the SUSY breaking scale is fixed for a given choice of $\mff$. On the other hand, for a scale $\mff$ fixed to be the inflaton mass $\tilde m$, it imposes an upper bound on the SUSY breaking scale $f$ (in Planck units)
\begin{equation}
f\ll \mff\,.
\end{equation} 

As a matter of fact, this result is model dependant and should be extended to a more extensive study of the possible supersymmetry breaking sector.

\section{Reheating}\label{Section:Reheating}

The reheating after inflation may be one of the most interesting features of the above scenario. In our model the right handed sneutrino is the inflaton and hence it can decay to MSSM particles via natural decay channels allowed by the gauge symmetry.  Here the situation is different from \cite{Heurtier:2015ima}, where it was necessary to add extra components to allow the inflaton to decay.

An implementation of the double seesaw mechanism, naturally appearing in flipped $SU(5)\times U(1)$ models \cite{Ellis:2011es}, gives mass to the right-handed neutrino upon integrating out the singlet fields $S$ and $\bar N^c$, of the order of $\mu_\phi$. In this context, successful inflation via right-handed sneutrino would imply that its mass is of order $\mu_\phi \sim {\cal O}(10^{12-13})$, the same as the right handed neutrino, since SUSY is restored at the end of inflation. Now the renormalizable superpotential that results in interactions between the inflaton $(\tilde{N^c})$ and the MSSM particles is given by
\bea\label{eq:Reheat_Superpot}
W_{SU(5)} &=&{ Y_u \,\mathbf{\bar{5}}_{H_u\alpha} \, \mathbf{10}_F^{\alpha\beta} \, \mathbf{\bar{5}}_{F\beta} } \, +\, Y_{d1}\, \epsilon_{\alpha\beta\gamma\delta\lambda}\, \mathbf{10}_F^{\alpha\beta}\, \mathbf{10}_F^{\gamma\delta}\, \mathbf{5}_{H_d}^{\lambda}\nonumber\\
& + & Y_{d2}\, \epsilon^{\alpha\beta\gamma\delta\lambda}\, \bar{\mathbf{10}}_{F\alpha\beta} \, \bar{\mathbf{10}}_{F\gamma\delta}\, \bar{\mathbf{5}}_{H_u \lambda}\, + \, Y_e \, \mathbf{5}_{H_d}^{\lambda} \, \mathbf{\bar{5}}_{F \lambda} \, \mathbf{1}_F\, \notag \\
W_{SO(10)} &=& Y_{\phi}\, \sigma^j_{\rho\eta} \mathbf{16}^\rho_F \mathbf{16}^\eta_F \mathbf{10}_{Fj} + Y_{\bar\phi} \, \bar\sigma_j^{\dot\rho \dot\eta}\bar{\mathbf{16}}_{F\dot\rho} \bar{\mathbf{16}}_{F\dot\eta} \mathbf{10}^j_F + Y_s \, \mathbf{10}^j_F\mathbf{10}_{Fj} \mathbf{1}_F.
\eea
where the indices $\alpha,\, \beta,\,... = 1,\dots,5$ are $SU(5)$ indices, both the vector indices $j = 1,\dots ,10$ and spinor indices $\rho,\eta,\dot\rho,\dot\eta = 1,\dots,16$ are $SO(10)$ indices, $\sigma$ and $\bar\sigma$ are the generalised Pauli matrices, and sum over flavour indices is implied.

The right handed neutrino superfield exists in the SM singlet component of $\mathbf{10}_F$ of $SU(5)$ and $\mathbf{16}_F$ of $SO(10)$. Therefore, for both cases, the inflaton can decay to the MSSM lighter fermions only via the following interaction Lagrangian
\be
{\cal L}_{int}=  -Y \, \tilde{N^c}\,\left( \nu_L \,\tilde{H}^0_u \, + \, e_L \,\tilde{H}^+_u\right),
\ee
which can be extracted from either of the first terms in eq.~\ref{eq:Reheat_Superpot} with $Y = Y_u$ or $Y_\phi$ accordingly. The reheating temperature $T_R$ is defined by~\cite{Lazarides:1996dv,Lazarides:2001zd}
\bea
T_R \approx \frac{ (8\pi)^{1/4}}{7}\left( \Gamma\,  M_p\right)^{1/2},
\eea
where $\Gamma$ is the total decay width of the inflaton field
\be
\Gamma = \Gamma_{\tilde{\nu}_{R} \to  \nu_L \,\tilde{H}^0_u} + \Gamma_{\tilde{\nu}_{R} \to  e_L \,\tilde{H}^+_u}\,,
\ee
and  $M_p$ is the reduced Planck mass. In this respect, the decay of the sneutrino to massless fermions is given by 
\be
\Gamma_{\tilde{\nu}_{R}}= \frac{|Y|^2 \mu_\phi}{8\pi}\,.
\ee

\begin{figure}[h]
 \centering
 \includegraphics[width=0.6\textwidth]{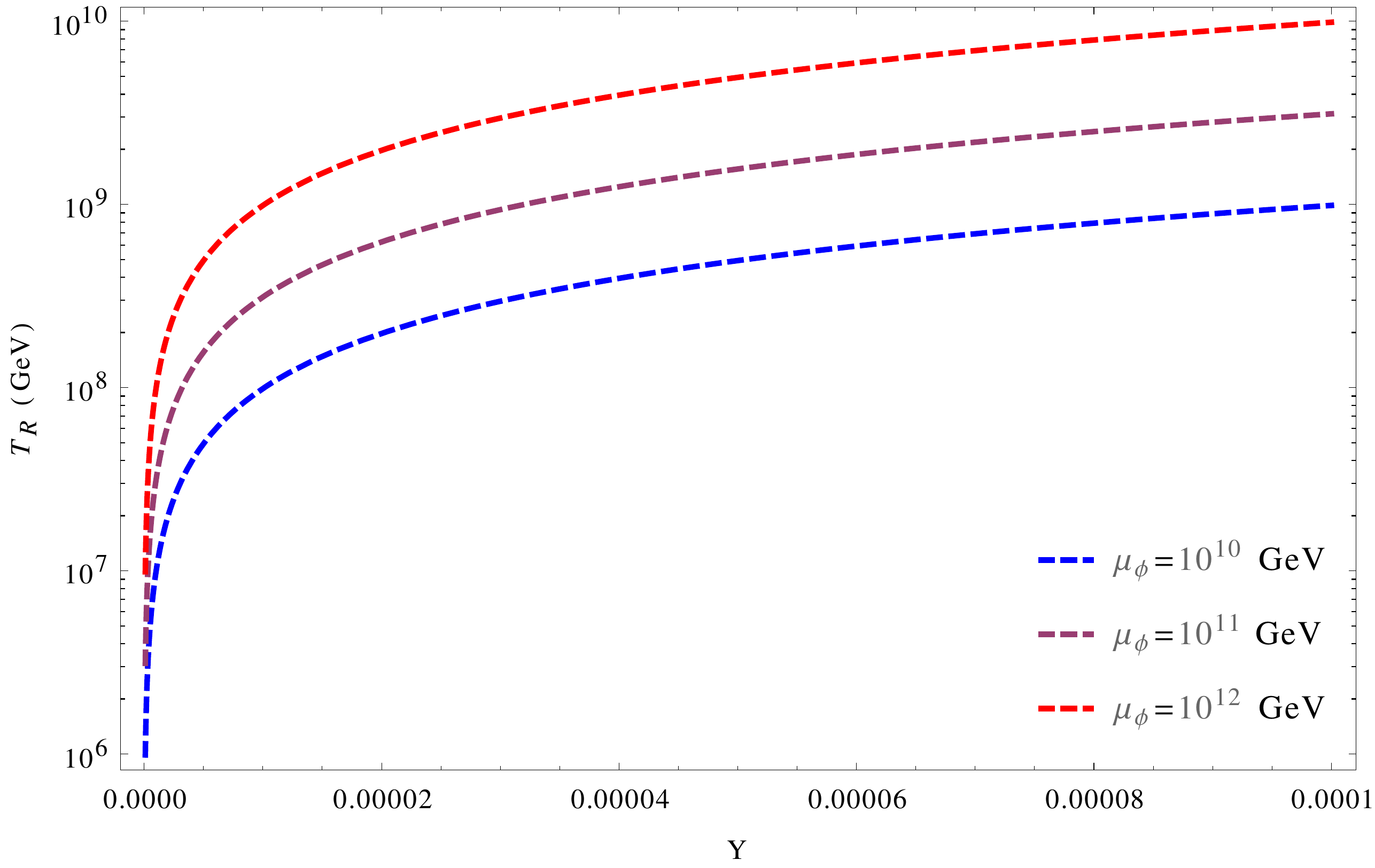}
 \caption{\footnotesize Reheating temperature $T_R$ versus the Yukawa coupling $Y$ for different values of the inflaton mass parameter $\mu_\phi = 10^{10}, 10^{11}, 10^{12}$ GeV.}
 \label{fig:reheating}
\end{figure}
Cosmological constraints such as the gravitino overproduction problem \cite{Ellis:1984eq, Ellis:1984er, Moroi:1993mb, Kawasaki:2004yh, Antusch:2010mv} impose an upper bound on the reheating temperature. This problem comes from the fact that if gravitinos are stable then their energy density should not be larger than the energy density of the universe, which results in an upper bound on the reheating temperature. However, unstable gravitinos can decay to the  lightest supersymmetric particle (LSP) if it is heavy enough, provided R-parity is conserved and, accordingly, the relic abundance of dark matter will contribute also to the bound on the reheating temperature. Taking into account that this decay may occur before, during and after the Big Bang Nucleosynthesis (BBN) in addition to the case of light gravitino, then  the decay of a light gravitino will contradict BBN observations. Therefore, due to the above cases, the constraint on the reheating temperature is given by \cite{Antusch:2010mv}
\begin{equation}
 T_R < 10^{7} - 10^{10} \text{ GeV}.
\end{equation}

Figure \ref{fig:reheating} shows the relation between the reheating temperature and the Yukawa coupling, for different values of the inflaton mass. For typical sneutrino masses of order $10^{12-13}$ GeV, the Yukawa couplings take the value $Y \lesssim 10^{-5}$ which is typically of order the up quark Yukawa\footnote{The up-quark sector masses are given by $Y v \sin \beta$. Since $10\lesssim\tan \beta \lesssim 50$ is favoured, as hinted by global fit scans such as \cite{deVries:2015hva}, therefore $ \sin \beta\sim {\cal O}(1)$.}. From this, we can conclude that the inflaton should correspond to sneutrino from the first generation in order to be consistent with the constraints on the reheating temperature. Other generations of sneutrinos, with larger Yukawa couplings, would potentially contribute significantly to the reheating temperature, driving it to very large values. In order to avoid this, we could assume the existence of a flavour violating sector at high energies that would stabilize the fields of the heavy generations prior to inflation, thus not affecting the reheating temperature. Specific details about this fall beyond the scope of this work.

\section{Conclusions and discussion}
\label{Section:Conclusions}

In this paper we have proposed a way to embed in supergravity a full set up of chaotic inflation within the framework of a GUT {\em flipped} group of the sort $\mathcal G \times U(1)_X$. The $\eta$-problem is evaded using the conjugate shift symmetry developed in \cite{Heurtier:2015ima} while the inflationary material is released in a similar fashion than in \cite{Ellis:2014xda}. Inflation is driven by the sneutrino whereas the spontaneous breaking of the GUT symmetry is triggered by a $U(1)_X$ charged heavy Higgs. The dynamics of all the fields is treated in full detail, including that of spectator fields, as well as the backreaction of the waterfall field on the inflationary potential.

Two scenarios of inflation are detailed: either (i) the critical point when the waterfall takes place is located at very low values of the inflaton and inflation is mostly released in the region where the waterfall field is stuck at the origin, or (ii) the waterfall happens at early stages of inflation, and the last 60 e-folds of inflation take place after the waterfall has happened. In the latter case, the inflaton direction goes along the minimum of the waterfall field.

Cosmological observables are computed in both scenarios and can both be accommodated to lay in the 2-$\sigma$ region of the last Planck measurements. However, we note that some fine tuning is necessary on the parameters in Eq. \eqref{finetuning} in order to have consistency with the observables, particularly in scenario (i) above. In the second case (ii) the breaking of the GUT symmetry takes place much earlier than inflation itself. Although this seems to suggest that the flipped structure of the unification is not needed anymore, we have seen that it remains actually crucial for releasing a satisfactory reheating period. In both scenarios we find that inflation and the observables follow from the value of $\mu_\phi$, as in standard chaotic inflation, which a priori has no relation to the symmetry breaking scale $M_{GUT}$. Though some artificial relation is imposed, through equations \eqref{eq:GUTbreaking}, \eqref{lhlf} and \eqref{finetuning}, we acknowledge that this is an artifice of our model and fine tuning, and hence this apparent relation should not be taken as a consequence of our analysis.

One should note that the question of supersymmetry breaking in the vacuum has not been addressed in this paper. However the presence of large F-term contributions in the vacuum could backreact on the inflation trajectory under the form of soft-terms in the scalar potential, as detailed in \cite{Buchmuller:2014pla, Buchmuller:2015oma}. Such implementation would strongly depend on the SUSY breaking sector added to the model. Simplest models would require a high SUSY breaking scale since the inflation sector is built from a quadratic term in the superpotential. However, models of sgoldstinoless inflation \cite{Ferrara:2014kva, DallAgata:2014qsj} could be accomodated in our model, rendereing low scale inflation possible to release. We have shown an example of the backreaction of SUSY breaking on the inflationary dynamics on a toy model, where we found a solution similar to the case with a stabilizer on the SUSY scale. More specific discussions are rather model dependent, so we defer them to future work.

Finally, the reheating temperature for these scenarios is computed as well and it turns out to be rather constraining. Indeed Yukawa couplings of order unity would lead to a unacceptably high reheating temperature and overclose the universe. The required value for the Yukawa coupling, $Y \sim 10^{-5}$, corresponds to that of the up quark, leading to the conclusion that the inflaton belongs to the first generation and, assumming the decoupling of the other two generations, could satisfy the constraints of reheating. This further motivates the choice of a flipped $SU(5)\times U(1)$ unification group, where the right-handed sneutrino has direct couplings to the rest of the MSSM particles.

\section*{Acknowledgements}
L.H. would like to thank W. Buchm\"uller for interesting discussions as well as the DESY Theory group of Hamburg for its hospitality during part of the realization of this work. The work of L.H. is funded by the Belgian Federal Science Policy through the Interuniversity Attraction Pole P7/37. T.E.G. was funded by the Research Council of Norway under FRIPRO project number 230546/F20. The work of A.M. is partially supported by the STDF project 13858 and the European Union's Horizon 2020 research and innovation programme under the Marie Skłodowska-Curie grant agreement No 690575.
\cleardoublepage
\bibliographystyle{h-physrev4}

\bibliography{bib}

\end{document}